\documentclass[letterpaper, 10 pt, conference]{ieeeconf}  
\usepackage{graphicx} 
\usepackage{svg}
\usepackage{cite}
\usepackage{amsmath,amssymb,amsfonts}
\usepackage{algorithmic}
\usepackage{textcomp}
\usepackage{xcolor}
\usepackage{booktabs}
\usepackage{algorithm}
\usepackage{url}

\IEEEoverridecommandlockouts                              

\title{\LARGE \bf
Personalized Sleep Prediction via Deep Adaptive Spatiotemporal Modeling and Sparse Data
}

\author{Xueyi Wang$^{1}$, C. J. C. (Claudine) Lamoth$^{2}$, and Elisabeth Wilhelm$^{3}$
\thanks{*This work is part of the project Healthy Living as a Service project (with the project number 13948 of the research programme KIC, which is (partly) financed by the Dutch Research Council (NWO)).}
\thanks{$^{1}$Xueyi Wang is with Engineering and Technology Institute Groningen, University of Groningen, Groningen, Netherlands.
        {\tt\small xueyi.wang@rug.nl}}%
\thanks{$^{2}$C. J. C. (Claudine) Lamoth is with the Department of Human Movement Sciences, University of Groningen, University Medical Center Groningen, Groningen, Netherlands.
        {\tt\small c.j.c.lamoth@umcg.nl}}%
\thanks{$^{3}$Elisabeth Wilhelm is with the Engineering and Technology Institute Groningen, University of Groningen,
        Groningen, Netherlands.
        {\tt\small e.wilhelm@rug.nl}}%
}

\begin{document}

\maketitle
\thispagestyle{empty}
\pagestyle{empty}

\begin{abstract}

A sleep forecast allows individuals and healthcare providers to anticipate and proactively address factors influencing restful rest, ultimately improving mental and physical well-being. This work presents an adaptive spatial and temporal model (AdaST-Sleep) for predicting sleep scores. Our proposed model combines convolutional layers to capture spatial feature interactions between multiple features and recurrent neural network layers to handle longer-term temporal health-related data. A domain classifier is further integrated to generalize across different subjects. We conducted several experiments using five input window sizes (3, 5, 7, 9, 11 days) and five predicting window sizes (1, 3, 5, 7, 9 days). Our approach consistently outperformed four baseline models, achieving its lowest RMSE (0.282) with a seven-day input window and a one-day predicting window. Moreover, the method maintained strong performance even when forecasting multiple days into the future, demonstrating its versatility for real-world applications. Visual comparisons reveal that the model accurately tracks both the overall sleep score level and daily fluctuations. These findings prove that the proposed framework provides a robust and adaptable solution for personalized sleep forecasting using sparse data from commercial wearable devices and domain adaptation techniques.
\newline

\indent \textit{Clinical relevance}— This is a method that could be used to inform participants and therapists in life-style interventions with the aim of improving sleep patterns about the participant's progress. 
\end{abstract}

\section{Introduction}
Sleep, a fundamental biological process, plays a vital role in human health, cognitive function, and overall well-being \cite{Rechtschaffen1989}. The ability to forecast sleep patterns and sleep quality has emerged as a critical area of research, with implications for both clinical applications and personal health management \cite{Zhang2021}. As global populations face increasing challenges related to stress, lifestyle-related non-communicable diseases , and environmental disruptions, developing robust sleep prediction models becomes increasingly important for individual and public health. 

Modern technological advancements, particularly in machine learning and wearable sensor technologies, have revolutionized our approach to sleep analysis. These innovations enable high-resolution, non-invasive monitoring of physiological parameters such as heart rate variability, and respiratory patterns~\cite{kleinsasser2022}. By leveraging sophisticated algorithmic techniques and wearable devices, researchers can now develop predictive models that not only detect sleep anomalies, but also forecast sleep quality parameters in the future.

\section{Related Work}


Traditional sleep studies relied heavily on polysomnography and clinical observations in sleep laboratories~\cite{raschella2022}. However, these methods prove impractical for real-world sleep forecasting due to the intrusive nature of monitoring equipment and the influence of unfamiliar sleep environments on natural sleep patterns \cite{kingshott2000effect}. Commercial devices like fitness tracker-activity watches provide a convenient way to monitor sleep quality and can offer valuable insights, although they are generally less accurate than polysomnography (PSG) \cite{stevens2019garmin}. The combination of commercial wearables and machine learning algorithms for sleep prediction has grown significantly, providing a more accessible and non-intrusive monitoring approach~\cite{kleinsasser2022}. Convolutional Neural Networks (CNNs) and Recurrent Neural Networks (RNNs) have been explored in extracting temporal features from sleep-related signals \cite{biswal2018expert}. However, traditional machine learning or deep learning based sleep forecasting models present several challenges, including individual variability in sleep patterns, the influence of external factors. 



The application of machine learning in sleep forecasting extends beyond clinical settings to personal health management. Smart devices and wearable technologies now incorporate sophisticated algorithms for sleep tracking and prediction, enabling users to optimize their sleep habits and identify potential sleep disorders early \cite{Patanaik2018}. These technological advancements have enabled extensive data collection and analysis, enhancing our ability to predict sleep patterns across diverse populations and understand how anticipated sleep behaviors correlate with various health outcomes \cite{zhang2022sleep}.

\begin{figure}
    \centering
    \includegraphics[width=1\linewidth]{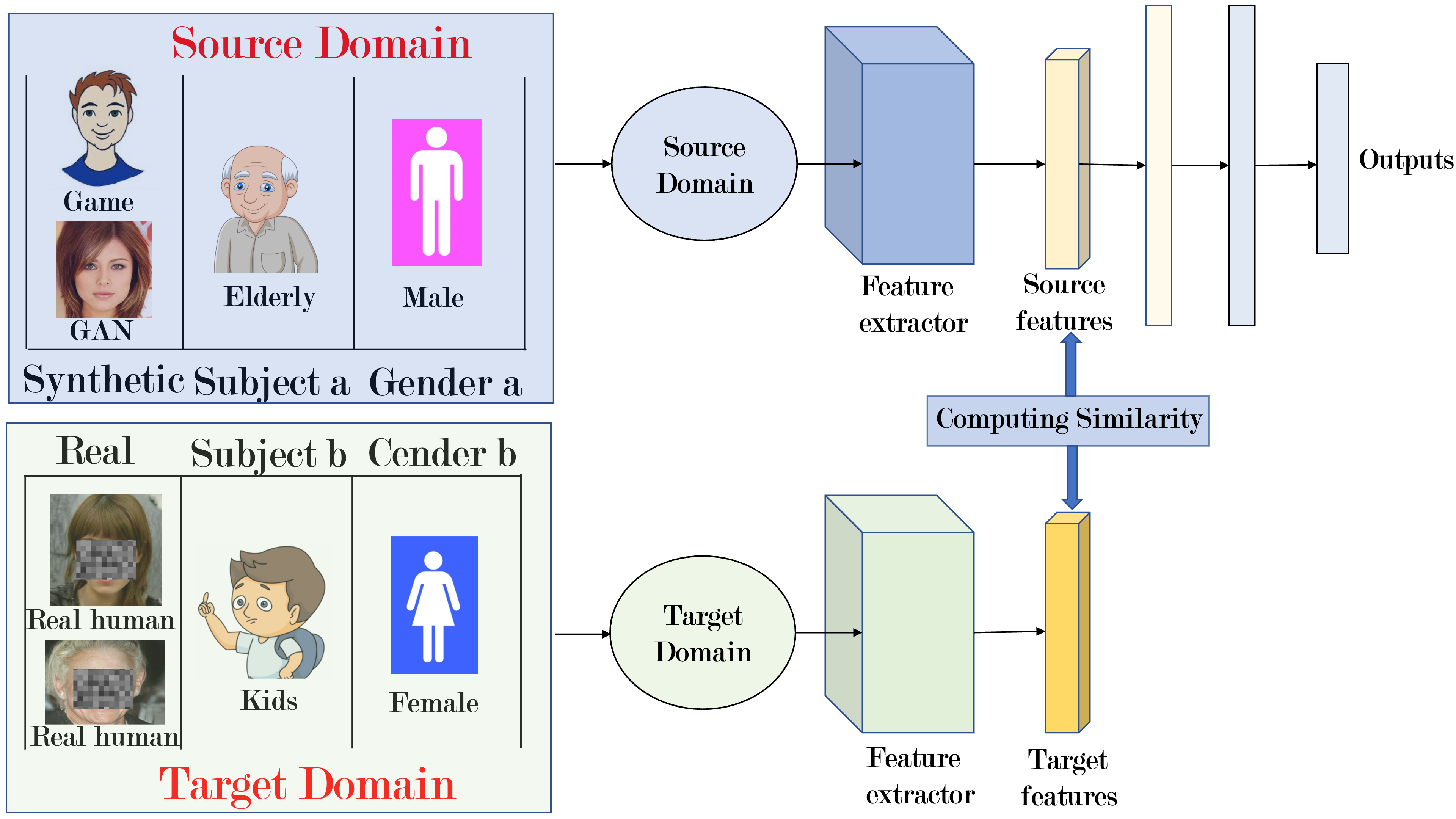}
    \caption{Illustration of domain adaptation for data from various subjects and conditions.}
    \label{fig:domain}
\end{figure}

Despite these advances, several challenges remain in sleep forecasting, including the generalization of sleep forecasting models across different subjects \cite{he2022single}. Traditional machine learning approaches often assume that training and test data are drawn from the same distribution. However, in sleep monitoring scenarios, this assumption rarely holds due to inter-subject variability and differences in data collection conditions \cite{tong2018sleep} as shown in Figure~\ref{fig:domain}. Individual variations in sleep patterns, physiological characteristics, and environmental factors often lead to performance degradation when models trained on one population are applied to new subjects \cite{Phan2021}. Domain adaptation techniques offer a promising solution to this challenge by enabling models to learn domain-invariant features while maintaining prediction accuracy. It addresses this limitation by explicitly modeling and minimizing the distribution shift between source and target domains \cite{zhao2021unsupervised}. We aim to address these limitations by proposing an adaptive spatiotemporal model to predict sleep scores from a sparse dataset in this work.

\section{Dataset}

We started by contributing a dataset (Wearlife-RUG) in this work based on  - Wearlife study (UMCG) from the Healthy Living as a Service project. It was conducted in accordance with the Declaration of Helsinki and Dutch regulations governing research involving human participants. The Central Ethics Review Board for non-WMO studies (CTc UMCG), which oversees investigations outside the scope of the Medical Research Involving Human Subjects Act (WMO), reviewed and approved the study (study register number 18021). Written informed consent was obtained from each participant prior to enrollment.

Seventeen participants initially began data collection using a smartwatch (Garmin Vivosmart 5). One withdrew after three weeks. Due to the short monitoring period, this subject was excluded from the analysis. The remaining 16 female participants had a median age of 55.5 years (interquartile range [IQR]: 16.25 years) and a median BMI of 28.22\,kg/m\(^2\) (IQR: 10.74\,kg/m\(^2\)). Seven of these participants had previously used a smartwatch. 


Participants were recruited from two separate field labs located in the northern region of the Netherlands. One field lab is a local walking group in a village in the Province of Groningen, where participants either receive referrals from healthcare providers or join on their own initiative. This group meets weekly for hour-long walks supervised by local healthcare professionals. The second field lab, located in a village in the Province of Drenthe, is a personal lifestyle intervention program. Participants enroll in a 10-week regimen featuring personalized nutrition plans, cooking workshops, small-group sports sessions twice a week, and lessons on mindfulness and healthy behavioral changes. For this study, individuals were eligible to participate if they were 18 years or older and able to speak and understand Dutch. The only exclusion criterion was the lack of a smartphone compatible with the Garmin Connect app, which is required for data recording. More information of the dataset can be found in the previous work \cite{wang2024multivariate}.

The Garmin Vivosmart 5 transmits sensor readings to the user’s phone via Bluetooth. Within the accompanying application, participants can view recorded data and health indicators, including heart rate, stress, and sleep scores. Low-resolution data (one data point per day) is stored in an online user profile under a pseudonymous username to maintain privacy. Only the birth year, weight, and height of the participant were provided, as these inputs are required by the smartwatch algorithms to translate sensor data into the health metrics presented. The research team downloaded data from this online interface on a weekly basis.

\begin{figure}
    \centering
    \includegraphics[width=1\linewidth]{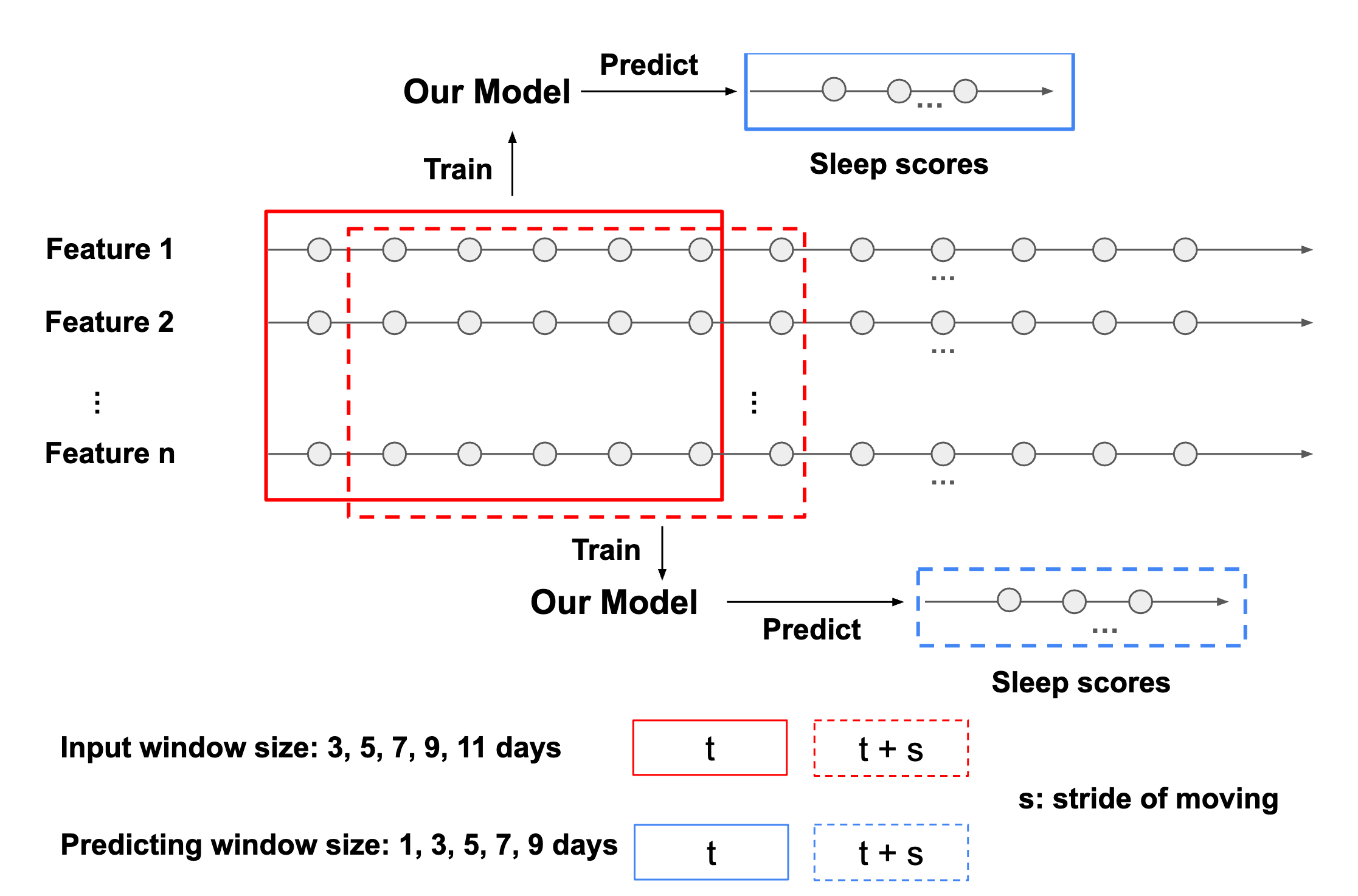}
    \caption{Sliding window for data augmentation with five input window sizes (3, 5, 7, 9, 11 days), five predicting window sizes (1, 3, 5, 7, 9 days) and the stride of one day.}
    \label{fig:sliding-window}
\end{figure}

All data are maintained in the secure, offline Virtual Research Workspace (VRW) administered by the University of Groningen. Prior to downloading and local analysis, data are preprocessed within the VRW in a secure, isolated environment to remove sensitive information. 

\section{Methods}
\subsection{Feature processing and selecting}
Missing entries can compromise analytical results and lead to misinterpretations. To address this concern, we used the SimpleImputer from sklearn to handle NaN and $-1$ obtained from the Garmin device. This replaces the missing value (NaN) with the mean of the corresponding column. The parameter of Hydration needed manual report in the app, but most subjects did not record that. Therefore, we removed the “Hydration” feature for all participants. In addition, the first and last days of each participant’s recordings have been excluded due to insufficient device usage on those specific days. For practical applications, this study employs a single daily data point (i.e., sparse data) rather than high-frequency measurements, significantly reducing storage and computational demands. Sleep is influenced by both daily activities and bodily signals~\cite{o'connor1995}. A total of 24 features from daily activity and body signals were used in this study, including total kilocalories, total steps, total distance, highly active seconds, active seconds, moderate-intensity minutes, resting heart rate, minimum average heart rate, maximum average heart rate, average waking respiration value, highest respiration value, lowest respiration value, stress average, deep sleep seconds, light sleep seconds, REM sleep seconds, awake sleep seconds, awake count, average sleep stress, restless moment count, lowest respiration, highest respiration, average respiration, and dates (working days or not). The complete explanation of Garmin sleep score can be found in \cite{australiagarmin}. In short it ranges from 0 to 100 and based on duration, quality, and heart rate variability, and it is excluded from features and used as the ground truth to train and forecast sleep.

\subsection{Sliding window}

\begin{figure*}
    \centering
    \includegraphics[width=0.82\linewidth]{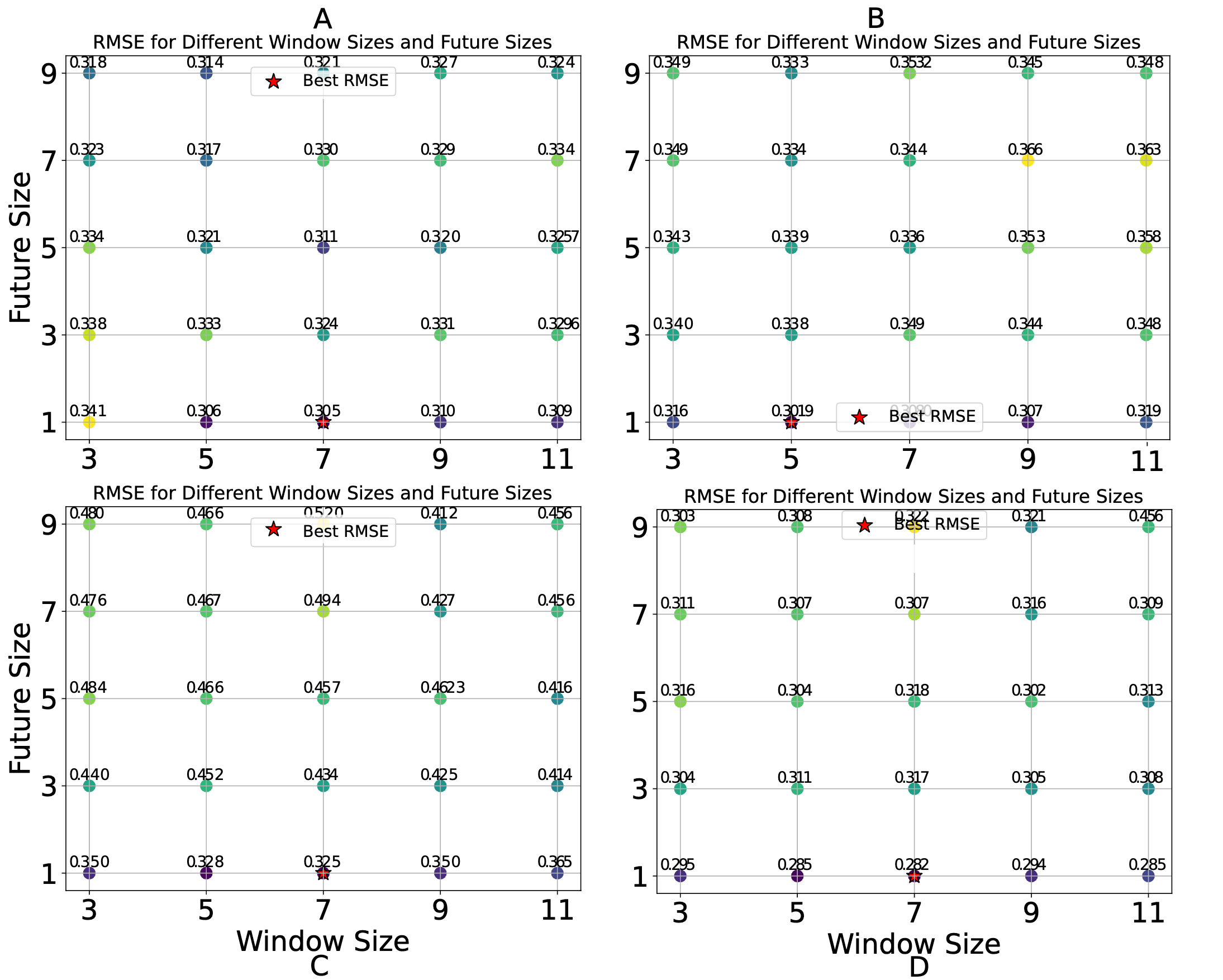}
    \caption{Averaged RMSE (all 16 subjects) of various input window sizes (3, 5, 7, 9, 11 days) and predicting window size (1, 3, 5, 7, 9 days) for four base models (A: BiLSTM, B: CNN, C: MLP, D; our model).}
    \label{fig:all-results}
\end{figure*}

In this work, we aim to predict sleep scores as defined by the Garmin, using multivariate data collected from Garmin watches. To pre-process the time series data, we employ a sliding window approach with window size (3, 5, 7, 9, 11 days), stride $s$ (one day), and prediction window size (1, 3, 5, 7, 9 days), transforming each subject's dataset into a set of windowed instances for data augmentation as shown in Figure~\ref{fig:sliding-window}.

\subsection{Evaluation metrics}
\noindent
We use Root Mean Square Error (RMSE): as the metric in this work due to its interpretability and computational simplicity. The definition of RMSE is given in equation \ref{eq}, where \(y_i\) denotes the actual normalized sleep score value on day \(i\), \(\hat{y}_i\) represents the corresponding predicted value, and \(n\) is the total number of days.
\begin{equation}
\text{RMSE} = \sqrt{\frac{1}{n} \sum_{i=1}^n (y_i - \hat{y}_i)^2}, \label{eq}
\end{equation}

\subsection{Leave one subject out cross validation}
In many machine learning problems involving human subjects, data collected from different individuals (subjects) may vary significantly due to physiological, behavioral, or demographic factors. One standard approach for evaluating the robustness of a model across different individuals is \emph{leave-one-subject-out cross-validation} (LOSO). This approach ensures that for each subject, the model is evaluated on data that remain completely unseen during training, providing a realistic estimate of cross-subject generalization performance. There are 16 distinct subjects in this work, each associated with a dataset:

\begin{equation}
   \mathcal{D}_i \;=\; \{(\mathbf{x}_{i,j}, \mathbf{y}_{i,j}) \;|\; j = 1, 2, \ldots, N_i\}
\end{equation}

where:
\begin{itemize}
    \item $\mathbf{x}_{i,j} \in \mathcal{X}$ is the feature vector of the $j$-th sample from subject $i$,
    \item $\mathbf{y}_{i,j} \in \mathcal{Y}$ is the value of sleep scores corresponding to $\mathbf{x}_{i,j}$,
\end{itemize}
Hence, the entire dataset by 16 subjects in this work can be written as

\begin{equation}
\mathcal{D} \;=\; \bigcup_{i=1}^{16} \mathcal{D}_i.
\end{equation}



In this work, we split the data by subject. Concretely, for each subject $i \in \{1, 2, \ldots, 16\}$:
All data were split the \emph{all data} (from the $N$ subjects) into a separate training set ($N - 2$ subjects), a validation set (one subject) and a test set (one subject). For instance,
\[
\text{Train set} = \mathcal{D}_p, \quad
\text{Validation set} = \mathcal{D}_q, \quad
\text{Test set} = \mathcal{D}_i,
\]
where $p \neq i$, $q \neq i$, and $p \neq q$. Hyperparameters are tuned on $\mathcal{D}_q$ (validation set), while ultimately evaluating performance on the unseen subject $\mathcal{D}_i$ (test set).


After performing this procedure for each subject $i$, we obtain a set of RMSE $\{RMSE_1, RMSE_2, \ldots, RMSE_i\}$, where $RMSE_i$ is, for example, the root mean squared error on the $i$-th fold. LOSO is a rigorous strategy for evaluating subject-based machine learning tasks~\cite{yin2017recognition}. By training on $N-2$ participants , validating on another participant and testing on the remaining participant, we obtain an unbiased estimate of cross-subject generalization. This guards against overfitting to subject-specific traits and provides a more reliable sense of how a model will generalize to entirely new individuals in real applications. 

\begin{figure}
    \centering
    \includegraphics[width=0.9\linewidth]{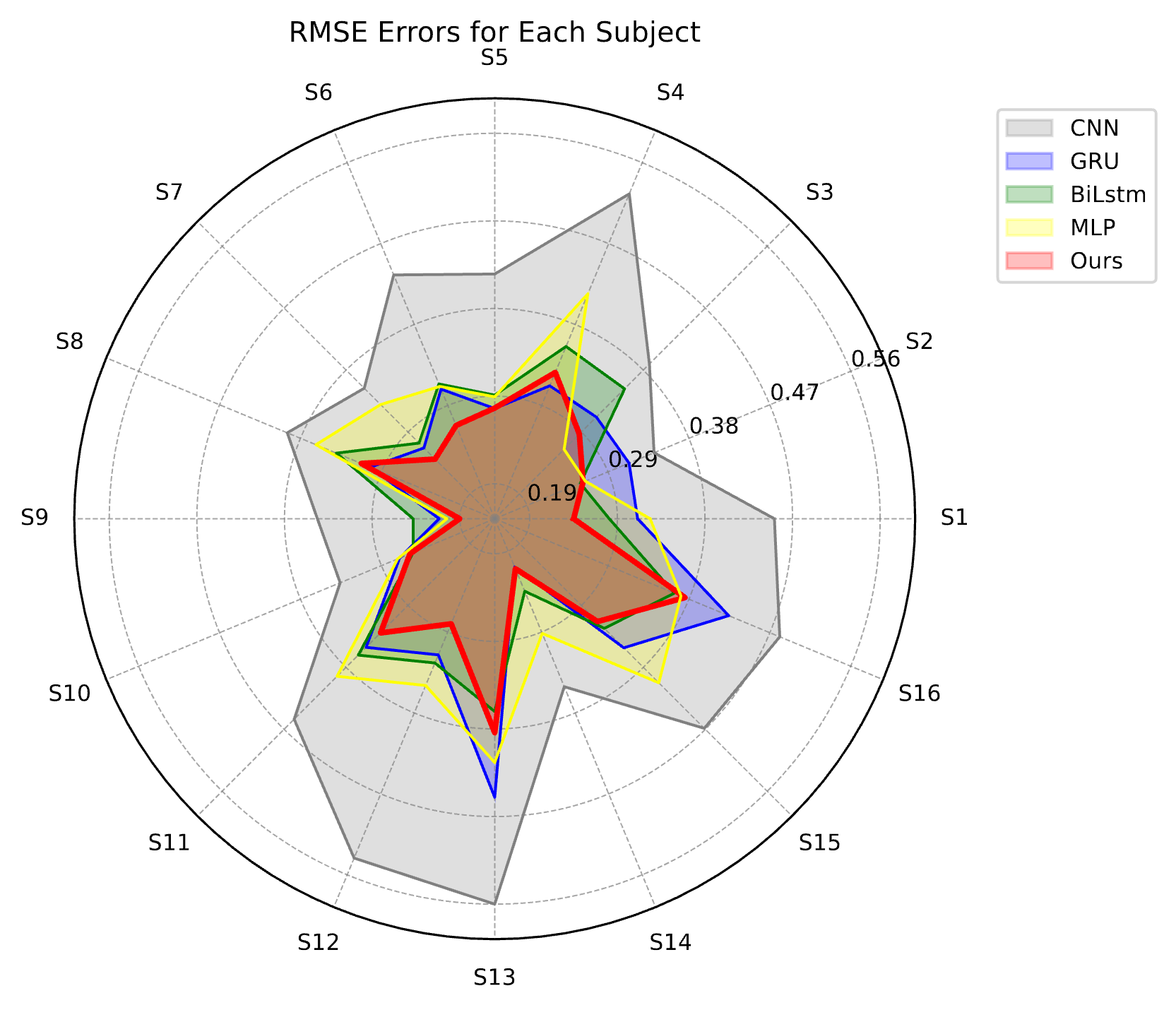}
    \caption{Comparison of RMSE for different models in terms of seven days input window size and one day predicting window size. Ours (AdaST-Sleep) refers to our adaptive spatial and temporal model.}
    \label{fig:radar-plot}
\end{figure}

\begin{figure*}
    \centering
    \includegraphics[width=0.72\linewidth]{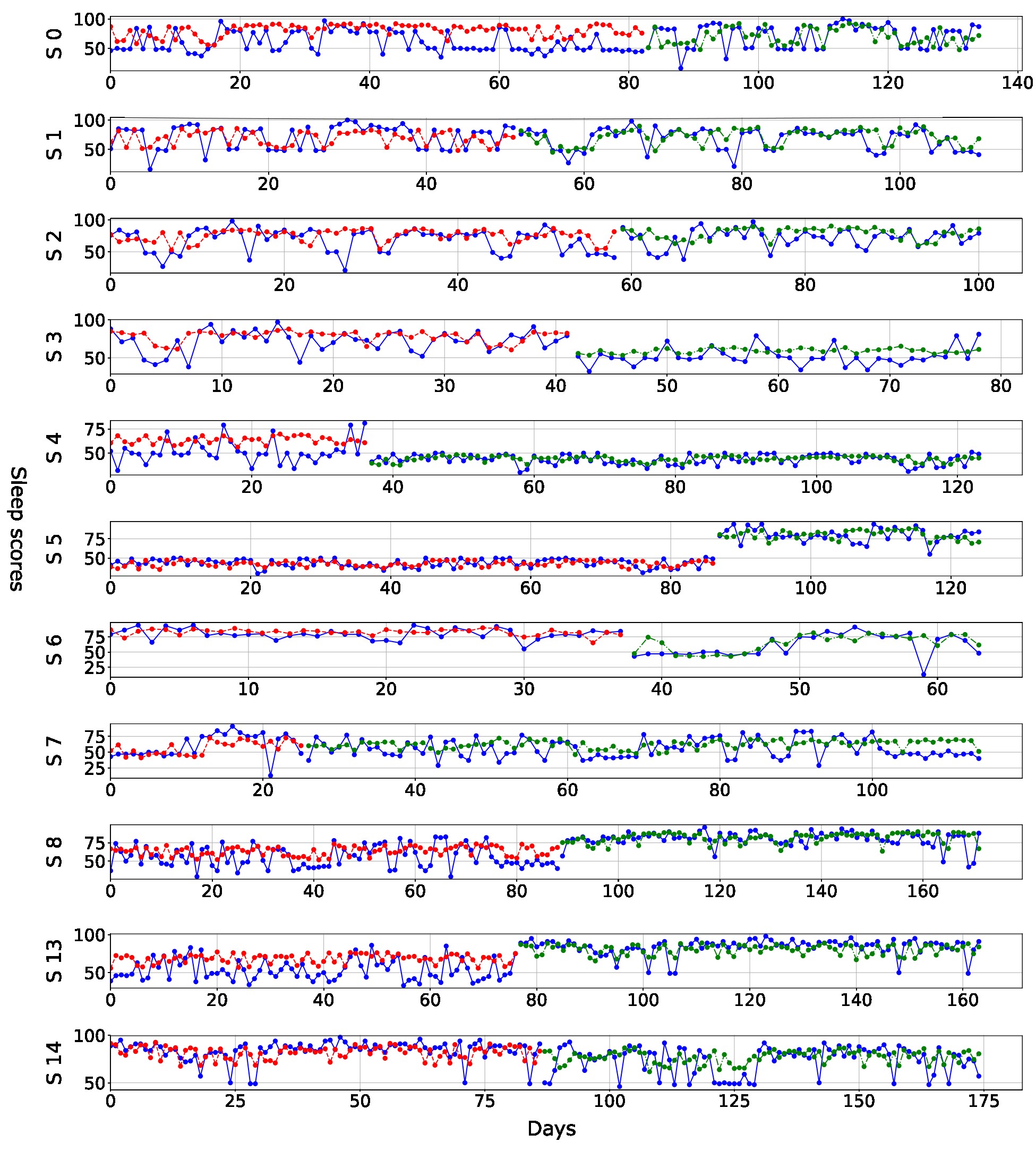}
    \caption{True values vs predictions in validation and test set for part of participants (blue lines: true values of sleep scores, red lines: prediction of validation set, green lines: predictions of test set). S-N denotes the number of subjects used in the test set. The x-axis represents days, and the y-axis represents sleep scores.}
    \label{fig:true-pre}
\end{figure*}

\subsection{Domain Adaptation}

In practical scenarios, the data distribution of the training set (source domain) may differ from that of the test set (target domain) due to inter-subject variability. This distributional shift can degrade the model's performance on unseen participants. $\mathcal{D}_s = \mathcal{D}_{\text{p}}$ represent the \emph{source domain} and $\mathcal{D}_t = \mathcal{D}_q$ denote the \emph{target domain} for a given fold $(i, q)$. The joint distributions of the source and target domains are:
\begin{equation}
    P_s(\mathbf{x}, \mathbf{y}) \; \neq \; P_t(\mathbf{x}, \mathbf{y})
\end{equation}

where $P_s$ and $P_t$ are the probability distributions over $\mathcal{X} \times \mathcal{Y}$ for the source and target domains, respectively.

\emph{Domain adaptation} seeks to mitigate this issue by aligning the source and target domains, thereby enhancing the model's generalization ability. When distributions vary significantly across subjects (due to physiology, behavior, etc.) as shown in Figure~\ref{fig:domain}, each subject can be viewed as its own \emph{domain}. Training on $\mathcal{D}_p$, validating on $\mathcal{D}_q$, and testing on $\mathcal{D}_i$ can be viewed as an adaptation process from a \emph{source domain} to a \emph{target domain}. This viewpoint motivates the application of \emph{domain adaptation} methods, which aim to reduce distributional discrepancies between training subjects and a new target subject. To enable domain adaptation, we include a domain\_classifier. It outputs classes for classification among possible domains.

\subsection{Models}
In this work, we have tested various base model, including Convolutional Neural Network (CNN), Bidirectional Long Short-Term Memory (BiLSTM), Gated Recurrent Unit (GRU) and Multilayer Perceptron (MLP). We also propose our own adaptive spatial and temporal model model, which leverages both convolutional neural networks (CNNs) and Recurrent Neural Network (RNNs), alongside a domain classifier for domain adaptation. 

Let $x \in \mathbb{R}^{B \times T \times F}$ represent a batch of input data, where $B$ is the batch size, $T$ is the temporal dimension (sequence length), and $F$ is the number of input features per time step. We first permute the input into the shape $[B, F, T]$ and feed it into a series of one-dimensional convolutions:

\begin{itemize}
    \item \textbf{First Conv Layer} (\texttt{conv1}) uses a kernel of size 3 with stride 1 and padding of 1, transforming $F$ input channels to \texttt{cnn\_hidden\_size} channels.
    \item An optional \textbf{Batch Normalization} (\texttt{bn1}) is applied if \texttt{use\_batchnorm} is set to \texttt{True}.
    \item A \textbf{ReLU} activation is then used, followed by a Dropout (\texttt{dropout1}) with rate \texttt{dropout\_cnn}.
\end{itemize}

When num\_conv\_layers is higher than two, we apply a second convolutional layer (\texttt{conv2}) similarly, potentially doubling the channel dimension to $2 \times \texttt{cnn\_hidden\_size}$, again with optional batch normalization, ReLU, and Dropout. The resulting output channels are then denoted \texttt{final\_channels}.

\begin{algorithm}
\caption{Train one model with domain adaptation}
\label{alg:train}
\begin{algorithmic}
\REQUIRE model $\mathcal{M}$, training loader $\mathcal{D}_\text{train}$, validation loader $\mathcal{D}_\text{val}$, epochs $E$, domain weight $\alpha$
\STATE Initialize optimizer (Adam) with $\text{lr} = 10^{-3}$
\FOR{$\text{epoch} = 1$ to $E$}
    \STATE $\mathcal{M}.\text{train}()$
    \FOR{each batch $(X, y, d) \in \mathcal{D}_\text{train}$}
        \STATE $\hat{y}, \hat{d} \leftarrow \mathcal{M}(X, \text{return\_domain=True})$
        \STATE $\mathcal{L}_\text{main} \leftarrow \text{RMSE}(\hat{y}, y)$
        \STATE $\mathcal{L}_\text{dom}  \leftarrow \text{CrossEntropy}(\hat{d}, d)$
        \STATE $\mathcal{L} \leftarrow \mathcal{L}_\text{main} + \alpha \cdot \mathcal{L}_\text{dom}$
        \STATE Backpropagate and update parameters
    \ENDFOR

    \STATE $\mathcal{M}.\text{eval}()$
    \STATE Evaluate $\mathcal{L}_\text{main}$ and $\mathcal{L}_\text{dom}$ on $\mathcal{D}_\text{val}$

    \STATE Print epoch results
\ENDFOR
\end{algorithmic}
\end{algorithm}

After the convolutional layers, the data is permuted back to shape $[B, T, \texttt{final\_channels}]$, and then passed to a multi-layer LSTM:
We use \texttt{num\_layers} stacked LSTM layers with a hidden dimension \texttt{lstm\_hidden\_size} and a dropout probability of \texttt{dropout\_lstm}. The LSTM processes each time step, returning hidden states for all time steps. We take the hidden state at the final time step, and feed it into a fully connected layer (\texttt{fc}) to predict the target.
This output corresponds to our main prediction objective.
\subsection{Training Procedure}

Our training loop (see Algorithm~\ref{alg:train}) uses two objectives:
\begin{itemize}
    \item A \textbf{main loss}, $\mathcal{L}_\text{main}$, typically RMSE for the model, comparing the predicted output $\mathbf{\hat{y}}$ with the ground-truth $\mathbf{y}$.
    \item A \textbf{domain classification loss}, $\mathcal{L}_\text{dom}$, which is a cross-entropy loss to distinguish domains $\mathbf{d}$.
\end{itemize}

We combine both losses into a single objective with a trade-off parameter $\alpha$:
\begin{equation}
   \mathcal{L} = \mathcal{L}_\text{main} + \alpha \cdot \mathcal{L}_\text{dom}
\end{equation}

A pseudo-code outline of the training process is given in algorithm \ref{alg:train}.

\subsection{Hyperparameter Fine-Tuning}
Hyperparameter tuning in this study is performed via Optuna, a framework that searches for optimal parameter values. Our tuning process focuses on key architectural and training hyperparameters, including: {num\_conv\_layers}, the number of convolutional layers (range: 1--2);        \texttt{num\_layers}, the number of LSTM layers (range: 1--3);
 \texttt{cnn\_hidden\_size}, the channel dimension in the CNN (choices: 16, 32, 64);
\texttt{lstm\_hidden\_size}, the LSTM hidden dimension (choices: 64, 128, 256);
\texttt{dropout\_cnn} and \texttt{dropout\_lstm}, the dropout rates in the CNN and LSTM, respectively (range: 0.1--0.5 in steps of 0.1);
\texttt{batch\_size} (choices: 8, 16, 32); and
\texttt{alpha}, the weight of the domain-adaptation loss term (range: 0.0--1.0 in steps of 0.1). All trials use an Adam optimizer with a fixed learning rate (\(\texttt{FIXED\_LR} = 1 \times 10^{-3}\)) and a weight decay of \(1\times 10^{-5}\). We run each study for \(\texttt{NR\_TRAILS} = 30\) trials or until a timeout occurs. Optuna logs the loss values and identifies the best-performing trial based on the lowest average RMSE over all participants. 

\subsection{Hardware and Software Setup}
\label{sec:hardware}

All experiments were conducted on two hardware platforms, a local workstation (Intel Core~i7-14700K CPU, an NVIDIA GeForce RTX~4090 GPU, and 32\,GB of RAM) and the H\'abr\'ok computing cluster\footnote{For more details, see \url{https://www.rug.nl/society-business/center-for-information-technology/research/services/hpc/habrok}.} (GPU-accelerated nodes featuring 64~CPU cores, 512\,GB of RAM, and four NVIDIA A100 GPUs). The H\'abr\'ok computing cluster is provided by the Center for Information Technology (CIT) at the University of Groningen. 
The model was primarily developed on the local workstation and tested on the GPU-accelerated nodes of H\'abr\'ok. All deep learning models in this work were implemented using PyTorch.

\section{Results and Discussion}

Overall, our experimental results confirm that the proposed adaptive spatial and temporal model effectively predicts sleep quality using a comprehensive set of 24 physiological and activity-related features. By combining convolutional layers to capture local feature interactions and LSTM layers to model extended temporal dependencies, and incorporating a domain classifier to adapt across different participants, the model consistently outperforms baseline approaches across various input and predicting window sizes.

\subsection{Evaluation with Different Window and predicting Sizes}
We tested five input window sizes (3, 5, 7, 9, and 11 days) in combination with five predicting window sizes (1, 3, 5, 7, 9 days) on both baseline models and our proposed adaptive spatial and temporal model. Figure~\ref{fig:all-results} summarizes the results. From this figure, it can be observed that all models, including our proposed approach, generally achieve their lowest RMSE (our model: 0.282) when the predicting window size is one day. Other models produced RMSE values ranging from 0.3047 to 0.4244. This is in line with many practical applications, as one-day forecasts can be beneficial for day to day interventions.

Sub-figure~(D) in Figure~\ref{fig:all-results} shows the outcomes of our proposed model. Notably, it outperforms the baseline models for all combinations of input and predicting window sizes, while the MLP baseline exhibits the highest RMSE in most scenarios. This discrepancy suggests that simple feed-forward networks may struggle to capture the complex and noisy patterns in the data, whereas our approach benefits from efficient spatial feature extraction and temporal modeling for different participants. 

Interestingly, as the input window increases beyond 7 days (e.g., to 9 or 11 days), the improvements become less pronounced, indicating a potential saturation where additional temporal context does not further reduce error. However, the model’s ability to maintain strong performance across longer predicting windows—achieving an RMSE of 0.303 for a 9-day prediction compared to a best RMSE of 0.282 for a 1-day prediction—highlights its versatility and potential usefulness in scenarios requiring multi-day forecasts.

Our model achieves its best performance with an input window size of seven days and a predicting window size of one day (RMSE of 0.282). Finally, Figure~\ref{fig:radar-plot} presents a radar plot comparing various baseline models and our proposed model across 16 participants in terms of seven days input window and one day predicting window. Each spoke in the radar plot corresponds to the RMSE for a specific participant. The shorter radial distances for our model indicate substantially lower error compared to the baselines in almost every case. This consistency suggests that the interaction of adaptive CNN, RNN-based temporal modeling, and domain adaptation techniques enables our approach to robustly handle participant-specific variability and produce accurate sleep forecasts.

\subsection{Visualization of True vs. Predicted Sleep Scores}
Figure~\ref{fig:true-pre} compares the ground-truth sleep scores against our model’s predictions on both the validation and unseen test sets. Several participants had a similar structure to others therefore we depict their data here as an example. One can observe that the model predicts not only the overall level of sleep scores, but also captures day-to-day changes in many participants. We employed a leave-one-subject-out cross-validation strategy, meaning each participant’s data is entirely excluded during training, then used for validation and testing. 

An example of data that is challenging for the classifiers is the data of subject 14. This contains multiple sudden drops and short term changes. Despite this shift, the model still provides reasonable predictions. This result highlights the importance of our domain-adaptive framework, which helps the model generalize to new, unseen subjects whose data characteristics may differ from those used in training.

Our model effectively captures trends and fluctuations in sleep scores, as shown for subjects 4 and 6 in Figure~\ref{fig:true-pre}. However, the model struggles to handle and predict values accurately when sudden anomalies and large changes in values occur, as seen in Subjects 1, 3, 6, 12, 14, and 15. 


\subsection{Limitations}
Our results demonstrate the feasibility of predicting sleep quality using 16 subjects, however, several limitations should be acknowledged. First, the monitoring duration in this study was relatively short, and data were recorded on a daily basis. Capturing data with higher time resolution (e.g., at minute-level intervals) and extending the overall monitoring period may better account for day-to-day variations and patterns. Second, although 16 subjects were sufficient to develop the core model, additional participants are needed to fully assess its generalization ability. 

\section{CONCLUSIONS}
In this paper, we introduced a new dataset (Wearlife-RUG), collected by  16 subjects in two towns of the Netherlands, and proposed an adaptive spatial and temporal model (AdaST-Sleep) for predicting sleep quality. Our approach combines convolutional layers for local CNN feature extraction, RNN layers for long-term temporal dependency modeling, and a domain classifier to handle subject-specific data shifts. Several experiments were conducted with varying input window sizes (3, 5, 7, 9, 11 days) and predicting window sizes (1, 3, 5, 7, 9 days), across multiple baseline models.

The results consistently show that our proposed model outperforms the baselines in terms of RMSE, particularly excelling with an input window size of seven days and a one-day predicting window. Furthermore, it maintains strong performance when forecasting for longer predicting windows, underscoring its adaptability to a range of temporal horizons. Qualitative visualizations on unseen subjects confirm that the model captures both the overall sleep score level and daily fluctuations, even when the data distributions differ between training and testing. By integrating efficient feature extraction, temporal modeling, and generalization to new subjects, the proposed framework shows promise for applications for life intervention where personalized sleep monitoring and forecasting are crucial.

\section*{ACKNOWLEDGMENT}
This work is part of the project Healthy Living as a Service project (with the project number 13948 of the research programme KIC, which is (partly) financed by the Dutch Research Council (NWO)). We would also like to thank the facilities provided by CIT of the University of Groningen, such as the cluster Habrok and other supports.

\bibliographystyle{unsrt}  
\bibliography{ref}         

\end{document}